
%
%
%
\input amssym.def
\input amssym.tex
\tolerance=2000\hbadness=2000
\overfullrule=0pt
\magnification=1200
\advance\hsize by 1.5truecm
\advance\hoffset by -1.15truecm
\advance\vsize by -0.75truecm
\advance\voffset by 1truecm
\parindent=0pt
\def\KK{I\!\!K}
\def\NN{I\!\!N}
\def\CC{C\!\!\!\!I}
\def\ZZ{Z\!\!\!Z}
\font\tenfrak=eufm10
\font\sevenfrak=eufm7
\font\fivefrak=eufm5
\newfam\frakfam
\textfont\frakfam=\tenfrak
\scriptscriptfont\frakfam=\fivefrak
\scriptfont\frakfam=\sevenfrak
\def\frak{\fam\frakfam\tenfrak}
\font\ninerm=cmr9
\font\bfeins=cmbx12
\font\bfzwei=cmbx12 scaled \magstep1

\def\abs{\par\vskip 0.3cm\goodbreak\noindent}
\def\Abs{\par\vskip 1.7cm\goodbreak\noindent}
\def\act{\triangleright}
\def\FF{\par\vfill\eject}
\def\IfFF{\FF\ifodd\pageno\else {\nopagenumbers{\centerline{}\eject}}\fi}
\def\IfPN{\ifnum\pageno=1\else
   {\hss{\ninerm ---  \quad\folio\quad  ---}\hss}\fi}
\def\litem{\par\noindent\hangindent=1.5cm\ltextindent}
\def\lfl{\leaders\hbox to 1em{\hss \hss}\hfill}
\def\nl{\par\noindent}

\def\Ues{\par\nobreak\vskip 0.75 cm\nobreak\noindent}

\def\Mittefrei#1#2{\hbox to \hsize{#1\hss#2}}
\def\ltextindent#1{\hbox to \hangindent{#1\hss}\ignorespaces}
\def\litem{\par\noindent\hangindent=1.5cm\ltextindent}
\def\3{\ss}
\def\ad{{\rm ad}}
\def\E{{\bf 1 \!\! {\rm l}}}
\def\fB{{\frak B}}
\def\fK{{\frak K}}
\def\fU{{\frak U}}
\def\fV{{\frak V}}
\def\fW{{\frak W}}
\def\fX{{\frak X}}
\def\fg{{\frak g}}

\def\fx{{\frak x}}
\def\fy{{\frak y}}

\def\id{{\rm id}}
\def\L#1#2#3{  {{L^{#1}}^{#2}}_{#3}  }
\def\ni{\hat{\cal O}}
\def\O#1#2#3#4{{{\Theta}{}^{#1#2}}_{#3#4}}
\def\t#1#2{  {t^{#1}}{}_{#2}  }
\def\X#1#2{{X^{#1}}_{#2}}

\footline={\IfPN}
\Mittefrei{September 1994}{UVA-FWI-Report 94-18}
\abs\abs
\vskip 1cm
\centerline{\bfzwei Braided Supersymmetry and (Co-)Homology}
\abs\abs\abs
\centerline{\bf Bernhard Drabant$^*$}
\abs\abs\abs
\centerline{\it Department of Mathematics and Computer Science}
\centerline{\it University of Amsterdam}
\centerline{\it Plantage Muidergracht 24, 1018 TV Amsterdam, Netherlands}
\centerline{\it email: {\rm drabant@fwi.uva.nl}}
\abs\abs\abs\abs
%
%
{\bf Abstract.}
\footnote{}{$^*${\ninerm Supported in part by the Deutsche
Forschungsgemeinschaft (DFG) through a research fellowship}}
Within the framework of braided or quasisymmetric monoidal categories
braided $\ni$-supersymmetry is investigated, where $\ni$ is a certain
functorial isomorphism $\ni : \otimes\cong \otimes$ in a braided symmetric
monoidal category. For an ordinary (co-)quasitriangular Hopf algebra
$(H,{\cal R})$ a braided monoidal category of $H$-(co-)modules with braiding
induced by the $\cal R$-matrix is considered.
It can be shown for a specific
class of $\ni$-supersymmetries in this category that every braided
$\ni$-super-Hopf algebra $\fB$ admits an ordinary $\ni$-super-Hopf algebra
structure on the cross product $\fB\rtimes H$ such that $H$ is a sub-Hopf
algebra and $\fB$ is a subalgebra in $\fB\rtimes H$. Applying these results to
the quantum Koszul complex $(K(q,\fg),{\rm d})$ of the quantum
enveloping algebra $U_q(\fg)$ for Lie algebras $\fg$
associated with the root systems $A_n$, $B_n$, $C_n$ and $D_n$ one
obtains a classical super-Hopf algebra structure on $(K(q,\fg),{\rm d})$
where the structure maps are morphisms of modules with differentiation.
\FF
%
%
{\bfeins 1. Introduction}
\Ues
In [DSO] a deformation of the (co-)homology of Lie algebras was found
for quantum enveloping algebras $U_q(\fg)$ of Lie algebras $\fg$ associated
with the root systems $A_n$, $B_n$, $C_n$ and $D_n$.
The deformed Kozul complex $(K(q,\fg),{\rm d})$ is the cross
product of $U_q(\fg)$ and the $q$-exterior algebra of forms $\Lambda(\fX)$
together with a derivative d obeying the usual properties. With regard
to the results of [Ma1, Ma2] the question now arises: Does the algebra
$K(q,\fg)$ admit a Hopf algebra structure (which respects the original
complex properties)? An answer to this question is given by the concept of
braided $\ni$-supersymmetry which can be considered as an extension of a
given braided symmetry and is itself a quasisymmetry --
braided symmetries were introduced in [JS].
It then turns out that the Koszul complex $(K(q,\fg),{\rm d})$
is a usual super-Hopf algebra [Man], and that the differentiation
d is compatible with this super-Hopf algebraic structure.
In some sense this ``dualizes'' the results of
M. Schlieker and B. Zumino who found that the algebra
of the bicovariant differential calculus on the quantum group $A_q(G)$ is a
super-Hopf algebra where the super-Hopf structure is respected by
the differentiation d [SZ].
Super-Hopf algebra structures on cross products of Hopf algebras
have also been found in [SWZ].
These examples are instructive applications of the more general
$\ni$-super-bosonization in certain braided $\ni$-supersymmetric
monoidal categories.
\nl
The paper is organized as follows. In Section 2 general facts
on braided symmetric monoidal categories
are reviewed [JS, Ma1] and the concept of braided $\ni$-supersymmetric
monoidal categories is introduced. By a braided $\ni$-supersymmetric
monoidal category we understand a braided symmetric monoidal category
with quasisymmetry $\Psi:\otimes\cong\otimes^\circ$ together with
a certain functorial isomorphism $\ni:\otimes\cong\otimes$ yielding
$\ni$-supersymmetry in the category.
$\ni$-Supersymmetric extensions of results of [Ma1, Ma2] are derived.
Especially $\ni$-super bosonization is investigated. In a category of
$H$-left modules of a quasitriangular Hopf algebra $(H,{\cal R})$ with
braided symmetry induced by the $\cal R$-matrix [Dri, Ma3],
$\ni$-super-bosonization for a specific class of functorial isomorphisms
will be carried out which converts any braided $\ni$-super-Hopf algebra
$\fB$ in this category to an ``ordinary'' $\ni$-super-Hopf algebra
$\fB\rtimes H$, which is the cross product of $\fB$ and $H$.
$\fB\rtimes H$ contains $\fB$ as subalgebra and $H$ as sub-Hopf algebra.
For the functorial isomorphism $\ni^{(1)}$ of
``classical'' supersymmetry these results are then used in
Section 3 to show that the deformed Koszul complex $(K(q,\fg),{\rm d})$ for Lie
algebras $\fg$ associated with the root systems $A_n$, $B_n$, $C_n$ and $D_n$
[DSO] is a super-Hopf subalgebra of the cross product of
the braided super-Hopf algebra $\Lambda(\fX)$ (in the category of
$\ZZ_2$-graded $D(U_q(\fg))$-modules) with the quantum double $D(U_q(\fg))$
of $U_q(\fg)$ (see also [SZ]). $\Lambda (\fX)$ is the algebra of
quantum exterior forms on the quantum group.
Furthermore it is derived that the original algebraic structure
of $(K(q,\fg),{\rm d})$ [DSO] enters into the super-Hopf algebra and that the
structure maps are morphisms of modules with differentiation [CE].
\abs
{\bf Acknowledgements:} I would like to thank T.H. Koornwinder and M. Schlieker
for stimulating discussions.
\Abs
%
%
{\bfeins 2. Braided Supersymmetry . . .}
\Ues
For categorical notations we refer to [Mac],
for the notion of braided monoidal categories, bosonization, transmutation,
braided Hopf algebras etc. we refer to [JS, Ma1, Ma2].
We begin with the definition of braided $\ni$-supersymmetry.
\abs
{\it Definition 2.1.} Let $({\cal C}, \otimes , \E ,\Psi )$ be a braided
monoidal category\footnote{$^\dagger$}{\ninerm For convenience we omit
throughout the paper the functorial isomorphisms which govern the associativity
of the tensor product and the unital property of the unit object.}
with bi-functor $\otimes$, unit object $\E$ and braided
symmetry $\Psi$, and let $\ni :\otimes\cong\otimes$ be a functorial isomorphism
such that
$$\ni\circ\Psi=\Psi\circ\ni\,,\eqno(2.1)$$
and
$$\eqalign{(\ni\Psi)_{\fU\,(\fV\otimes\fW)} &=
  (\id_{\fV}\otimes\ni_{\fW\,\fU})\circ\Psi_{\fU\,(\fV\otimes\fW)}\circ
  (\ni_{\fU\,\fV}\otimes\id_{\fW})\,,\cr
  (\ni\Psi)_{(\fU\otimes\fV)\,\fW} &=
  (\ni_{\fW\,\fU}\otimes\id_{\fV})\circ\Psi_{(\fU\otimes\fV)\,\fW}\circ
  (\id_{\fU}\otimes\ni_{\fV\,\fW})\cr
  \forall\ \fU,\,\fV,\,\fW &\in {\rm Ob}({\cal C})\,.\cr}
\eqno(2.2)
$$
Then $({\cal C}, \otimes , \E ,\ni\Psi )$ is called
{\it braided $\ni$-supersymmetric monoidal category} w.r.t.
the braided symmetry $\Psi$.
If $\Psi^2= \id$ then we speak of an $\ni$-supersymmetric monoidal category
w.r.t. $\Psi$.
\abs
{\it Remark.} The id-supersymmetry w.r.t. $\Psi$ is just $\Psi$ itself.
\abs
The following proposition shows that the unit object is ``bosonic'' and
that $\ni\Psi$ is again a braiding such that (2.1) holds. This is
equivalent to Definition 2.1.
\abs
{\bf Proposition 2.2.} $\ni\Psi$ is a braided symmetry, i.e.
$({\cal C}, \otimes , \E ,\ni\Psi )$ is a braided tensor category and
$$\ni_{\fX\,\E}=\id_{\fX\otimes\E}\,,\ \
  \ni_{\E\,\fX}=\id_{\E\otimes\fX}\quad
  \forall\ \fX \in {\rm Ob}({\cal C})\,.
\eqno(2.3)
$$
\abs
{\it Proof.} $\ni\Psi :\otimes\cong\otimes^{\circ}$
is a functorial isomorphism and the further properties of a braided
symmetry can be derived simply by applying Definition 2.1.
Since $\ni\Psi$ and $\Psi$ are braidings we obtain
$(\ni\Psi)_{\fX\,\E}=\ni_{\E\,\fX}\,\Psi_{\fX\,\E}=\Psi_{\fX\,\E}$. This
completes the proof.\lfl$\square$
\abs
In the braided monoidal category $({\cal C}, \otimes , \E ,\ni\Psi )$
we can define algebras, coalgebras, bi- and Hopf algebras w.r.t. $\Psi$.
We call them {\it braided $\ni$-super-bi-, braided $\ni$-super-Hopf algebras}
etc. If it is clear from the context we omit the addendum ``w.r.t. $\Psi$''.
\abs
{\bf Lemma 2.3.} Let $(\fB, \eta_{\fB}, {\rm m}_{\fB})$ and
$(\fK, \varepsilon_{\fK}, \Delta_{\fK})$ be an algebra and a coalgebra
in the braided $\ni$-supersymmetric monoidal
category $({\cal C}, \otimes , \E ,\ni\Psi )$ respectively. Then
$\forall\ \fV\in {\rm Ob}({\cal C})$
$$\eqalign{\ni_{\fB\,\fV}(\eta_{\fB}\otimes\id_\fV) &=
           \eta_{\fB}\otimes\id_\fV\,,\cr
           \ni_{\fV\,\fB}(\id_\fV\otimes\eta_{\fB}) &=
           \id_\fV\otimes\eta_{\fB}\cr}
\eqno(2.4)
$$
and
$$ \eqalign{(\varepsilon_{\fK}\otimes\id_{\fV})\,\ni_{\fK\,\fV} &=
            (\varepsilon_{\fK}\otimes\id_{\fV})\,,\cr
            (\id_{\fV}\otimes\varepsilon_{\fK})\,\ni_{\fV\,\fK} &=
            (\id_{\fV}\otimes\varepsilon_{\fK})\,.\cr}
\eqno(2.5)
$$
I.e. the unit and the counit are ``bosonic''.
\abs
{\it Proof.} Since $\ni$ is a functorial isomorphism and $\eta_{\fB}$,
$\varepsilon_{\fK}$ and $\id_{\fV}$ are morphisms in the category,
the lemma can be proved using Proposition 2.2.\lfl$\square$
\abs
{}From now on we restrict our considerations to monoidal categories
$({\cal M}, \otimes, \KK)$ where $\KK$ is a field,
the objects and morphisms in $\cal M$ are in particular
$\KK$-vector spaces and $\KK$-vector space homomorphisms,
and the usual tensor transposition $\tau$ is a symmetry in $\cal M$.
Let $(H,{\cal R})$ be a quasitriangular Hopf algebra in
$({\cal M},\otimes ,\KK ,\tau)$ and let
$({}_H{\cal M}, \otimes, \KK)$ be a monoidal category of
$H$-left modules and $H$-left module morphisms in $\cal M$ containing
$H$ as a module. In ${}_H{\cal M}$ the braided symmetry $\Psi_0$ induced by the
$\cal R$-matrix [Dri, Ma3] is supposed to exist.
$$\Psi_{0\,\fU\fV}(u\otimes v) := \sum {\cal R}_2\act v\otimes {\cal R}_1
  \act u
\eqno(2.6)
$$
where $\fU,\,\fV \in {\rm Ob}({}_H{\cal M})$, $u\in\fU$, $v\in\fV$,
${\cal R}=\sum {\cal R}_1\otimes {\cal R}_2 \in H\otimes H$ and $\act$ is
the action of $H$ on $\fU$ and $\fV$
respectively\footnote{${}^\dagger$}{\ninerm
In the dual language of coquasitriangular Hopf algebras and comodules
[Ma1, Wei] the definitions and results can be formulated similarly.}.
Now we investigate a certain class of functorial isomorphisms
$\ni :\otimes\cong\otimes$ with the following properties.
\abs
{\it Definition 2.4.}
\nl
i. $\ \ ({\cal M}, \otimes, \KK, \ni\tau)$ and
$({}_H{\cal M}, \otimes, \KK, \ni\Psi_0)$
are braided $\ni$-supersymmetric monoidal categories.
\nl
ii. $\ \ni_{H\,\fV}= \id_{H\otimes\fV}$ and $\ni_{\fV\,H}=
\id_{\fV\otimes H}$ $\forall\,\fV\in {}_H{\cal M}$, i.e. $H$ is ``bosonic''.
\nl
iii. $[(\id_{\fU}\otimes h\,\act ),\,\ni_{\fU\,\fV}]=0$ and
$[(h\act \otimes \,\id_{\fV} ),\,\ni_{\fU\,\fV}]=0$ $\forall\,\fU,\,\fV\in
{}_H{\cal M}$, $h\in H$.
\abs
Then for these supersymmetries the $\ni$-bosonization theorem can be stated.
\abs
{\bf Theorem 2.5.} Let $\fK$ be a braided $\ni$-super-Hopf algebra
in the category ${}_H{\cal M}$. Then the space $\fK\otimes H$ can be
equipped with an $\ni$-super-Hopf algebra structure in the category
$\cal M$, which is the cross product of $\fK$ and $H$ and is denoted by
$\fK\rtimes_{\ni} H$. Explicitely $\fK\rtimes_{\ni} H$ has the structure
$$\eqalign{\hat\eta &= \eta_{\fK}\otimes\eta_H\,,\cr
           \hat m((u\otimes a)\otimes (v\otimes b))
           &= \sum u\,(a_{(1)}\act v)\otimes a{(2)}\,b\,,\cr
           \hat\varepsilon &= \varepsilon_{\fK}\otimes\varepsilon_H\,,\cr
           \hat\Delta (u\otimes a)
           &= \sum u_{(1)}\otimes {\cal R}_{(2)}a_{(1)}\otimes
           {\cal R}_1\act u_{(2)}\otimes a_{(2)}\,,\cr
           \hat S(u\otimes a)
           &= \sum (S_H({\cal R}\,a)_{(1)}{\cal R}_1)\act S_{\fK}(u)
           \otimes S_H({\cal R}\,a)_{(2)}\cr}
\eqno(2.7)
$$
where $u\otimes a,\,v\otimes b\in \fK\otimes H$, $\Delta_H(a)=\sum a_{(1)}
\otimes a_{(2)}$ $\forall\ a\in H$
and $\Delta_\fK(u)=\sum u_{(1)}\otimes u_{(2)}$ $\forall\ u\in \fK$.
\abs
{\it Proof.} The definitions in (2.7) yield morphisms in $\cal M$.
It is verified immediately that $\hat\varepsilon$ is an algebra morphism.
The proofs of the algebra and coalgebra properties of $\fK\rtimes_{\ni} H$ and
the proof of the identities
$(\id\otimes\hat S )\circ\hat\Delta = (\hat S\otimes\id )\circ\Delta =
\hat\eta\circ\hat\varepsilon $ follow rather analogously like the corresponding
proofs in [Ma1, Wei]. It remains to show that $\hat\Delta$ is an algebra
morphism in the category $\cal M$. An easy calculation shows that
$\hat\Delta (\hat\E)= \hat\E\otimes\hat\E$. We sketch the proof of the
multiplicativity of $\hat\Delta$. One observes that
$$(\ni\tau)_{(\fK\otimes H)\,(\fK\otimes H)} =
  (\id_{\fK}\otimes\tau_{(\fK\otimes H)\,H})\circ (\ni_{\fK\,\fK}\otimes
  \id_{H\otimes H})\circ (\tau_{(\fK\otimes H)\,\fK}\otimes\id_H)\,.
  \eqno(2.8)
$$
Equation (2.8) can be obtained by taking into account that $\ni$ and $\tau$
fulfill Definition 2.4. With the notation $\ni_{\fK\,\fK}(u\otimes v)=:\sum
u^\uparrow\otimes v^\uparrow $ one gets on the one side
$$\eqalignno{&\hat\Delta\circ\hat m \left((u\otimes a)\otimes (v\otimes b)
           \right)&\cr
          =&\sum u_{(1)}\,
          \left(({\cal R}_2\,a_{(1)})\act v_{(1)}\right)^\uparrow\otimes
          {\cal R}_{2'}a_{(3)}b_{(1)}\otimes
          {\cal R}_{1'}\act\left(({\cal R}_1\act u{(2)})^\uparrow(a_{(2)}\act
          v_{(2)})\right)\otimes a_{(4)}b_{(2)}&(2.9)\cr
          =&\sum u_{(1)}\,(({\cal R}_2 a_{(1)})\act v_{(1)})^\uparrow\otimes
          {\cal R}_{2'}{\cal R}_{2''} a_{(3)} b{(1)}\otimes
          (({\cal R}_{1'}{\cal R}_1)\act u_{(2)})^\uparrow\,
          ({\cal R}_{1''} a_{(2)})\act v_{(2)}\otimes a_{(4)} b_{(2)}\,.&\cr}
$$
On the other side one obtains
$$\eqalignno{&(\hat m\otimes\hat m)\,(\id_{\fK\otimes H}\otimes
  (\ni\tau)_{(\fK\otimes H)\,(\fK\otimes H)}\otimes \id_{\fK\otimes H})
  (\hat\Delta\otimes\hat\Delta)\left((u\otimes a)\otimes(v\otimes b)\right)
  &(2.10)\cr
  =&\sum u_{(1)}\left(({\cal R}_{2\,(1)} a_{(1)})\act v_{(1)}\right)^{
  \uparrow}\otimes
  {\cal R}_{2\,(2)} a_{(2)}{\cal R}_{2'} b_{(1)}\otimes
  ({\cal R}_1\act u_{(2)})^\uparrow (a_{(3)}{\cal R}_{1'})\act v_{(2)}\otimes
  a_{(4)} b_{(2)}&\cr
  =&\sum u_{(1)}\,(({\cal R}_2 a_{(1)})\act v_{(1)})^\uparrow\otimes
  {\cal R}_{2'}{\cal R}_{2''} a_{(3)} b{(1)}\otimes
  (({\cal R}_{1'}{\cal R}_1)\act u_{(2)})^\uparrow\,
  ({\cal R}_{1''} a_{(2)})\act v_{(2)}\otimes a_{(4)} b_{(2)}\,.&\cr}
$$
In both cases we used the fact that $H$ is a quasitriangular Hopf algebra and
$\fK$ is a braided Hopf algebra in the category
$({}_H{\cal M},\otimes,\KK,\ni\Psi_0)$. Comparing the two results yields the
statement and thus Theorem 2.5 is proved.\lfl $\square$
\abs
A straightforward consequence of Theorem 2.5 is the following corollary.
\abs
{\bf Corollary 2.6.} In $\fK\rtimes_{\ni} H$ the Hopf algebra $H$ is embedded
through the Hopf algebra isomorphism
$H\cong \E_\fK\otimes H\subset\fK\rtimes_{\ni} H$, and $\fK$ considered as an
algebra is embedded through the algebra isomorphism
$\fK\cong \fK\otimes \E_H\subset\fK\rtimes_{\ni} H$.\lfl$\square$
\abs
Theorem 2.5 can be formulated for weaker conditions than those
listed in Definition 2.4. Let $({\cal M},\otimes,\KK,\tau^d)$ be a braided
monoidal category which contains $H$ as a (quasitriangular) Hopf algebra,
and let $({}_H{\cal M},\otimes,\KK,\Psi_0^d)$ be a braided monoidal category
of $H$-left modules in $\cal M$ containing $H$ as canonical module.
Assume that there exist two further monoidal categories
$({\cal M}^\circ,\otimes,\KK)$, and $({}_H{\cal M}^\circ,\otimes,\KK)$ which
consists of $H$-left modules in ${\cal M}^\circ$,
and two forgetful monoidal functors $U$ and ${}_HU$ such that the diagramm
{\input amstex
\def\diag#1#2#3#4#5#6#7#8{\CD                        
                         {#1} @> {#5} >> {#2} \\     
                         @V {#6} VV @V {#8} VV \\    
                         {#4} @> {#7} >> {#3}        
                         \endCD}                     
$${\diag {_H\Cal M}{_H\Cal M^\circ}{\Cal M^\circ}{\Cal M}
      {_HU}{V}{U}{V^\circ}}
\tag{D-1}$$
is} commutative, where the forgetful functors $V$ and $V^\circ$ are
canonical. Suppose that for any $\fU,\,\fV\in {\rm Ob}({}_H{\cal M})$
a morphism $\ni_{\fU\fV}: \fU\otimes\fV\to \fU\otimes\fV$ in
${\cal M}^\circ$ exists and let the usual tensor transposition
$\tau_{\fU\fV}$ be a morphism in ${\cal M}^\circ$ and the $\cal R$-matrix
induced transposition $\Psi_{0\,\fU\fV}$ according to eq.$\,\,$(2.6) be a
morphism in ${}_H{\cal M}^\circ$ such that
\nl
i. $\ \ \tau^d_{\fU\fV}= \ni_{\fV\fU}\circ\tau_{\fU\fV}$ and
   $\Psi^d_{0\,\fU\fV}= \ni_{\fV\fU}\circ\Psi_{0\,\fU\fV}$.
\nl
ii. $\ \ni_{H\,\fV}= \id_{H\otimes\fV}$ and $\ni_{\fV\,H}= \id_{\fV\otimes H}$.
\nl
iii. $[(\id_{\fU}\otimes h\,\act ),\,\ni_{\fU\,\fV}]=0$ and
     $[(h\act \otimes \,\id_{\fV} ),\,\ni_{\fU\,\fV}]=0$.
\nl
Then a theorem analogous to Theorem 2.5 holds and a counterpart of
Corollary 2.6 can be deduced similarly. An example for this construction
which is {\it not} a braided $\ni$-supersymmetry is provided by the category
of complexes [Par, Ma1]. The category ${\cal M}:=\KK$-Comp$_2$ with objects
being complexes which have a canonical $\ZZ_2$-grading and with
morphisms being $\ZZ_2$-graded homomorphisms of modules with differentiation,
i.e. ${\rm d}\circ f = f\circ {\rm d}$, admits a braided symmetry
$$\tau^d_{\fU\fV}(u\otimes v)= (-1)^{\hat u\hat v}\,v\otimes u +
  \lambda\,(-1)^{(\hat u -1)\,\hat v}\,{\rm d}_{\fV}(v)\otimes {\rm d}_{\fU}(u)
  \eqno(2.11)
$$
where $\fU,\,\fV\in {\rm Ob}({\cal M})$, $u\in \fU$, $v\in \fV$ are homogeneous
elements of degree $\hat u,\,\hat v\in\ZZ$ respectively, $\lambda\in\KK$ and
d$_\fU$, d$_\fV$ are the corresponding derivatives. The tensor derivative is
given through
${\rm d}_{\fU\otimes\fV}={\rm d}_\fU\otimes\id_\fV+\gamma_\fU\otimes{\rm
d}_\fV$
with grade indicating vector space homomorphism
$\gamma_\fU (u):= (-1)^{\hat u} u$ (see Proposition 3.1). In the category
${}_H{\cal M}:= H$-$\KK$-Comp$_2$ the objects being the
$H$-left modules in $\cal M$ where the derivative d is $H$-module morphism
and the action is $\ZZ_2$-graded, the morphisms in ${}_H{\cal M}$ are
$\ZZ_2$-graded $H$-module morphisms of modules with differentiation.
There exists a braiding $\Psi^d_0$ in ${}_H{\cal M}$.
$$\Psi^d_{0\,\fU\fV}(u\otimes v)= \sum (-1)^{\hat u\hat v}\,
  {\cal R}_2\act v\otimes {\cal R}_1\act u +
  \lambda\,(-1)^{(\hat u -1)\,\hat v}\,
  {\cal R}_2\act {\rm d}_{\fV}(v)\otimes {\cal R}_1\act {\rm d}_{\fU}(u)
  \eqno(2.12)
$$
where $\fU,\,\fV\in {\rm Ob}({}_H{\cal M})$ and $u\in \fU$, $v\in \fV$ are
homogeneous elements of degree $\hat u,\,\hat v\in\ZZ$ respectively. Let
${\cal M}^\circ$ be the category of $\ZZ_2$-graded $\KK$-vector spaces
and let ${}_H{\cal M}^\circ$ be the category of $\ZZ_2$-graded
$H$-left modules in ${\cal M}^\circ$. Then for all objects
$\fU,\,\fV\in {\cal M}$ a morphism $\ni_{\fU\fV}:\fU\otimes\fV\to\fU\otimes\fV$
in ${\cal M}^\circ$ can be defined.
$$\ni_{\fU\fV}(u\otimes v)= (-1)^{\hat u\hat v}\,u\otimes v +
 \lambda\,(-1)^{\hat u\,(\hat v-1)}\,{\rm d}_{\fU}(u)\otimes {\rm d}_{\fV}(v)
  \eqno(2.13)
$$
where $\fU,\,\fV\in {\rm Ob}({\cal M})$ and $u\in \fU$, $v\in \fV$ are
homogeneous elements of degree $\hat u,\,\hat v\in\ZZ$ respectively.
In this setting the above mentioned extension of Theorem 2.5 and of Corollary
2.6 applies. The notation is compatible. However there exists no braided
$\ni$-supersymmetry since the morphisms $\ni_{\fU\fV}$ in eq.$\,\,$(2.13)
do not induce a functorial morphism
$\ni : \otimes\buildrel {{}_\bullet}\over\to \otimes$.
Explicit examples of braided $\ni$-supersymmetries are given at the
end of Section 2.
\abs
For simplicity let in the following $({\cal M},\otimes ,\KK ,\tau)$ be
the monoidal category of $\NN_0$-graded $\KK$-vector spaces and
$({}_H{\cal M},\otimes ,\KK ,\Psi_0)$ be the monoidal category of
$\NN_0$-graded $H$-left modules in $\cal M$.
The quasitrangular Hopf algebra $(H,{\cal R})$ is an object in $\cal M$
through the identification $H=H_0$. The $H$-module structure on $H$ is
the canonical one. In these categories we consider $\ni$-supersymmetries
according to Definition 2.4.
\abs
{\bf Lemma 2.7.}  [Ma2, Wei]
For any $\fX \in {\rm Ob}({}_H{\cal M})$ the tensor space
$\fX^{\otimes n}$ is an $H$-left module in ${}_H{\cal M}$ and the
tensor algebra $T(\fX)=\bigoplus_{n=0}^\infty \fX^{\otimes n}$ is an algebra
in the category ${}_H{\cal M}$ with the usual tensor multiplication $m_T$,
unit $\E_T=1\in\KK$ and module action according to
$$\eqalign{h\act \E_T &:= \varepsilon_H(h)\,\E_T\,,\cr
           h\act (x_1\otimes\ldots\otimes x_n) &:=
           \sum h_{(1)}\act x_1\otimes\ldots\otimes h_{(n)}\act x_n\,,\cr
    h\act \sum_{n=0}^\infty \fx_n &:= \sum_{n=0}^\infty h\act\fx_n\cr}
\eqno(2.14)
$$
where $h\in H$, $x_i\in \fX$ $\forall\,i\in \{1,\ldots ,n\}$,
$\fx_n \in \fX^{\otimes n}$ $\forall\,n\in \{0,1,2,\ldots\}$.\lfl $\square$
\abs
For the functorial isomorphisms $\ni$ under consideration we suppose henceforth
that for any $\fV\in {\rm Ob}({}_H{\cal M})$ it holds
$$\ni_{T(\fX)\,\fV} = \bigoplus_{n=0}^\infty \ni_{\fX^{\otimes n}\,\fV}\,,\quad
  \ni_{\fV\,T(\fX)} = \bigoplus_{n=0}^\infty \ni_{\fV\,\fX^{\otimes n}}\,.
\eqno(2.15)
$$
Then we obtain the following proposition which can be proved in the same way
as the corresponding statement in [Ma2, Wei] exploiting strongly the
braided symmetry properties of $\ni\Psi_0$.
\abs
{\bf Proposition 2.8.} $T(\fX)$ is a braided $\ni$-super-Hopf algebra in
the category $({}_H{\cal M},\otimes,\KK,\ni\Psi_0)$ with the algebra structure
like in Lemma 2.7. Comultiplication, counit and antipode are defined
for any $x\in\fX$ through
$$\Delta_T (x)= x\otimes \E_T + \E_T\otimes x\,,\quad
           \varepsilon_T(x) = 0\,,\quad S_T(x) = -x
\eqno(2.16)
$$
and by multiplicative continuation according to
$$\eqalign{
  \Delta_T(\fx\,\fy)&= (\id_T\otimes(\ni\Psi_0)_{T\,T}\otimes\id_T)
  (\Delta_T(\fx)\otimes\Delta_T(\fy))\,,\cr
  \varepsilon_T(\fx\,\fy) &=\varepsilon_T(\fx)\,\varepsilon_T(\fy)\,,\cr
  S_T(\fx\,\fy)&= m_T\,(\ni\Psi_0)_{T\,T}(S_T(\fx)\otimes S_T(\fy))\cr
  \forall\ \fx,\,\fy &\in T(\fX)\cr}
\eqno(2.17)
$$
for the whole algebra $T(\fX)$.\lfl $\square$
\abs
Assume furthermore that $(\ni\Psi_0)_{\fX\,\fX}$ decomposes into projectors
as follows.
$$(\ni\Psi_0)_{\fX\,\fX} = \sum_{i=1}^n \alpha_i\, P_i\,,\quad
   P_i\,P_j = \delta_{ij}\,P_j\,,\quad
   \sum_{i=1}^n P_i = \id_{\fX\otimes\fX}
\eqno(2.18)
$$
where $i,j\in\{1,\ldots ,n\}$, $\alpha_i \in \KK$, $\alpha_i \ne \alpha_j$ for
$i\ne j$. Then each projector $P_i$ is a morphism in ${}_H{\cal M}$. It is easy
to see that for any $i\in\{ 1,\ldots ,n\}$ the ideal $J_i$ generated by
$m_T\,P_i\,(\fX\otimes\fX )$ is an object in ${}_H{\cal M}$ which does not
contain $\E_T$ and is invariant under $\ni\Psi_0$, i.e.
$(\ni\Psi_0)_{T\,\fV}(J_i\otimes \fV) \subset \fV\otimes J_i$ and
$(\ni\Psi_0)_{\fV\,T}(\fV\otimes J_i) \subset J_i\otimes \fV$
$\forall\,\fV\in {\rm Ob}({}_H{\cal M})$. Therefore
$$\bar T^i := T(\fX)\big/J_i\quad\forall\ i\in \{1,\ldots ,n\}\eqno(2.19)$$
is an $H$-module algebra in ${}_H{\cal M}$ and $\overline{\ni\Psi_0}$ can
be defined canonically for $\bar T^i$. Suppose now that
$$\overline{(\ni\Psi_0)_{T\,\fV}} = (\ni\Psi_0)_{\bar T^i\,\bar \fV}\,,
  \quad
  \overline{(\ni\Psi_0)_{\fV\,T}} = (\ni\Psi_0)_{\bar \fV\,\bar T^i}
\eqno(2.20)
$$
for any $i\in\{ 1,\ldots ,n\}$ and any $\fV\in {\rm Ob}({}_H{\cal M})$.
Then one finds like in [Ma2, Wei]
\abs
{\bf Proposition 2.9.} If there exists an $i_0\in \{1,\ldots ,n\}$ such that
$\alpha_{i_0}= -1$ in the projector decomposition (2.18), then $\bar T^{i_0}$
is a braided $\ni$-super-Hopf algebra in the category
$({}_H{\cal M},\otimes ,\KK ,\ni\Psi_0 )$. The structure maps of $\bar T^{i_0}$
are canonically induced by the corresponding maps of $T(\fX)$.\lfl $\square$
\abs
{\it Remark.} Through the assignment
$$\fU\times \fV\mapsto \ni^{(\alpha )}_{\fU\,\fV}\,,\quad
  \ni^{(\alpha )}_{\fU\,\fV} :
  \cases{\fU\otimes \fV &$\to \fU\otimes\fV$\cr
   u_i\otimes v_j &$\mapsto (-\alpha)^{ij} u_i\otimes v_j$\cr}
\eqno(2.21)
$$
where $\alpha\in \KK{\setminus}\{0\}$, $\fU,\,\fV\in {\rm Ob}({}_H{\cal M})$,
$u_i\in \fU_i$, $v_j\in \fV_j$, $i,\,j\in \NN_0$, a functorial isomorphism
${\ni^{(\alpha)} : \otimes\cong\otimes}$ is
defined on $({\cal M}, \otimes, \KK )$ which is nice enough to fulfill
{\it all} the properties supposed in this
section\footnote{${}^\dagger $}{\ninerm This
functorial isomorphism also appeares in [Ko1, Ma4].}. Thus if one
identifies an arbitrary $H$-left module $\fX$
with $\fX\cong\fX_1$ then $\fX\in {\rm Ob}({}_H{\cal M})$ and
$$(\ni^{(\alpha)}\Psi_0)_{\fX\fX} = (-\alpha)\,\Psi_{0\,\fX\fX}\eqno(2.22)$$
which simply rescales the braiding. In the context of [Ma5, SWW, Wei] rescaling
yields a new rescaling generator to obtain an ordinary Hopf algebra
while in the context of $\ni$-supersymmetry rescaling yields
$\ni$-super-Hopf structures. This fact will be used in Section 3
in the case of ``classical'' supersymmetry, i.e. where $\alpha =1$,
to obtain the desired results (see also [SZ]). When we speak henceforth of
``braided super-...'' without any further indication, we are working with
braided $\ni^{(1)}$-supersymmetry.
\Abs
%
%
{\bfeins 3. . . . and (Co-)Homology}
\Ues
The results of Section 2 are now applied to the complex
$(K(q,\fg),{\rm d})$ which is a deformation of the Koszul complex of
Lie algebras $\fg$ associated with the root systems $A_n$, $B_n$, $C_n$ or
$D_n$. Here we suppose $q\ne {\rm ``root}$ of unity''. We begin
by recalling the most important results of [DSO]. It is known that the
adjoint representation $\ad : U_q(\fg)\otimes \fX\to\fX$ of the quantum
enveloping algebra $U_q(\fg)$ in the $\CC$-vector space $\fX$ of
$U_q(\fg)$-generating vector fields  is defined through
$$\ad_u(x)= \sum u_{(1)}\,x\,S_U(u_{(2)})\quad\forall\ x\in\fX\,,u\in
   U_q(\fg)
  \eqno(3.1)
$$
where $S_U$ is the antipode of $U_q(\fg)$.
The vector space $\fX$ corresponds to the right invariant
vector fields associated with a
certain bicovariant differential calculus on quantized simple
Lie groups, such that $\fX$ generates $U_q(\fg)$ [CSWW, DJSWZ, Jur, Wor].
A $\CC$-basis in $\fX$ is given through [DSO]
$$\X ab = \delta^a_b\,\E_{U}- \sum_{k=1}^N \L +ak
  S_{U}(\L -kb)
\eqno(3.2)
$$
where $N= n+1$ for $A_n$, $N=2n+1$ for $B_n$ and $N=2n$ for $C_n$ and $D_n$,
$a,\,b\in \{1,\ldots ,N\}$, and $(\L \pm rs)_{r,s=1,\ldots ,N}$ are the
regular functionals of the corresponding quantum group [FRT]. The space $\fX$
is
dual to the space $\Gamma_{inv}$ of right invariant one-forms [Wor]
with basis $\left\{\eta_b^{\ a}|a,\,b \in \{1,\ldots ,N\}\right\}$ and with
left adjoint coaction $\Phi_\Gamma :\Gamma_{inv}\to A_q\otimes \Gamma_{inv}$,
where $A_q$ is the quantum group dual to $U_q(\fg)$. It holds [Jur, Wor]
$$\eqalign{<\X ab , \eta_c^{\ d}&> = \delta^{ad}_{\ \ cb}\,,\cr
      <\Phi_\Gamma (\eta), u\otimes x &> = <\eta, \ad_u (x)>\,,\cr
      \Phi_\Gamma (\eta_r^{\ s}) = \sum_{t,u} &S_{A}(\t ur)\,\t st\otimes
      \eta_u^{\ t}\cr}
\eqno(3.3)
$$
where $\eta\in \Gamma_{inv}$ and $u\otimes x \in U_q(\fg)\otimes \fX$. The
action of $\fX$ on $\fX$ can also be written as a deformed Lie bracket [Wor].
$$\ad_x(y)= x\,y - m_{U}\,\sigma (x\otimes y)
\eqno(3.4)
$$
where $x,\,y\in\fX\subset U_q(\fg)$, $m_U$ is the multiplication in $U_q(\fg)$
and $\sigma :\fX\otimes\fX \to \fX\otimes \fX $ is a linear transformation with
very specific properties [DSO, Jur, Wor]. For the vector fields under
consideration, $\sigma$ can be written as a projector decomposition
$$\sigma =\sum _{\rho\in R} \rho\, P_\rho\,, \quad
  P_\rho \,P_{\rho '}= \delta_{\rho\rho '} \,P_\rho\quad {\rm and}\quad
  \sum _{\rho\in R} P_\rho =\id_{\fX\otimes \fX}\,.
\eqno(3.5)
$$
where $\rho,\rho'\in R$. The set $R $ consists of complex numbers and it
contains $1$. The space $\Lambda(\fX)$ is given by
$$\Lambda(\fX) = T(\fX)\big/ (m_T P_1 (\fX\otimes \fX))
  = \bigoplus_{j=0}^\infty \Lambda_j (\fX)\,.
\eqno(3.6)
$$
It becomes an $\NN_0$-graded and thus $\ZZ_2$-graded $U_q(\fg)$-module
algebra through the action $\ad^\wedge$ which is induced by $\ad$ and
therefore
$$K(q,\fg) := \Lambda(\fX) \,\,{}^{\ad^\wedge}\!\!\!\!{\rtimes}\, U_q(\fg)
\eqno(3.7)
$$
is an $\NN_0$-graded algebra which contains $\Lambda (\fX)$ and $U_q(\fg)$ as
subalgebras [DSO]. Both in $\Lambda (\fX)$ and in $U_q(\fg)$ the vector fields
$\fX$ are contained. For distinction it is written $\fX \subset\Lambda (\fX)$
and $\tilde\fX \subset U_q(\fg)$. On $K(q,\fg) = \bigoplus_{j=0}^\infty
\Lambda_j(\fX)\,U_q(\fg) = \bigoplus_{j=0}^\infty K_j(q,\fg)$ a grade
indicating algebra isomorphism $\gamma: K(q,\fg)\to K(q,\fg)$ exists
such that $\gamma (a_j) = (-1)^j a_j$ $\forall\,a_j\in K_j(q,\fg)$. The
central theorem of [DSO] states that on $K(q,\fg)$ a derivative d
with the following properties can be uniquely defined.
$$\eqalign{ &{\rm d\ is\ }U_q(\fg){-}{\rm module\ morphism},\cr
            &{\rm d}(x) = \tilde x \in U_q(\fg)\quad \forall\ x\in
             \Lambda (\fX)\,,\cr
            &{\rm d}(k\,l) = {\rm d}(k)\,l + \gamma(k)\,{\rm d}(l)\quad
             \forall\ k,\,l \in K(q,\fg)\,.\cr}
\eqno(3.8)
$$
{}From this it follows that ${\rm d}^2 =0$ and
${\rm d} \gamma + \gamma {\rm d} =0$. The complex $(K(q,\fg), {\rm d})$ is a
deformation of the Koszul complex of the Lie algebras $\fg$ associated with
the root systems $A_n$, $B_n$, $C_n$ and $D_n$. We also need a
chain complex structure on the tensor product
$K^\otimes(q,\fg) := K(q,\fg)\otimes K(q,\fg)=
\bigoplus_{j=0}^\infty K^\otimes_j(q,\fg)$ where
$K^\otimes_j(q,\fg) = \bigoplus_{n+m=j} K_n(q,\fg)\otimes K_m(q,\fg)$. This is
guaranteed by the next proposition.
\abs
{\bf Proposition 3.1.} The tensor space $K^\otimes (q,\fg)$ is a super
complex algebra with multiplication map ${m^\otimes = (m_K\otimes m_K)\circ
(\id_K\otimes (\ni^{(1)}\tau )_{K\,K}\otimes \id_K)}$, with grade indicating
algebra isomorphism $\Gamma = \gamma\otimes\gamma$, and with derivative
${\rm D}= ({\rm d}\otimes \id_K) + (\gamma\otimes {\rm d})$ such that
$$\eqalign{{\rm D}(K^\otimes_j(q,\fg)) &\subset K^\otimes_{j-1}(q,\fg)\,,\cr
           {\rm D}(u\,v) = {\rm D}(u)\,&v + \Gamma(u)\,{\rm D}(v)\,,\cr
           {\rm D}^2 &= 0\,,\cr
           {\rm D} \Gamma + &\Gamma {\rm D} = 0\cr}
\eqno(3.9)
$$
where $j\in\NN_0$, $u,\,v\in K^\otimes (q,\fg)$ and $u_j\in
K^\otimes_j(q,\fg)$.
\abs
{\it Proof.} All the statements can be checked directly on homogeneous
elements using the properties of d and $\gamma$. After linear continuation
the statements follow.\lfl $\square$
\abs
The quantum enveloping algebra $U_q(\fg)$ is a quasitriangular Hopf algebra
with $\cal R$-matrix $\cal R$ [Dri, Ma3, Res]\footnote{$^\dagger$}
{\ninerm The matrix $\cal R$ is an element of a certain
completion of the tensor product of $U_q(\fg)$ [Dri, Ma3, Res]. In this more
general setting of quasitriangular Hopf algebras the notations and
results especially of Section 2 remain unchanged.}.
Similar as in [SZ] a representation of the quantum double
$D(U_q(\fg))$ of $U_q(\fg)$ [CEJSZ, Dri] in the space $\Lambda (\fX)$ will
be constructed. Here the $\cal R$-matrix
$${{\cal R}^\circ}= \sum {{\cal R}_2}^{-1}\otimes {\cal R}_{1'}
  \otimes \Delta_{U}({{\cal R}_1}^{-1}{\cal R}_{2'})
\eqno(3.10)
$$
will be used which makes $D(U_q(\fg))$ a quasitriangular Hopf algebra.
The structure maps are defined through
$$\eqalign{m^\circ &= (m_U\otimes m_U)\circ (\id_U\otimes \tau_{U\,U}\otimes
             \id_U)\,,\cr
           \eta^\circ &= \eta_U\otimes \eta_U\,,\cr
           \Delta^\circ &= {\cal
R}^{-1}{}_{23}\,(\id_U\otimes\tau_{U\,U}\otimes
            \id_U)\circ (\Delta_U\otimes\Delta_U)(\,.\,)\,{\cal R}_{23}\,,\cr
           \varepsilon^\circ &= \varepsilon_U\otimes\varepsilon_U\,,\cr
    S^\circ &= {\cal R}_{21}\, (S_U\otimes S_U)(\,.\,)\,{\cal
R}^{-1}{}_{21}\cr}
\eqno(3.11)
$$
with obvious notation [FRT]. We consider the mapping
$$\Phi^\circ_\Gamma : \cases{\Gamma_{inv} &$\to D(A_q)\otimes \Gamma_{inv}$\cr
             \eta_r^{\ s} &$\mapsto \sum_{t,u} S_A(\t ur)\otimes\t st
             \otimes \eta_u^{\ t}$\cr}
\eqno(3.12)
$$
where $D(A_q)$ is the dual quantum double of $D(U_q(\fg))$ [CEJSZ, Dri].
This defines a left coaction of $D(A_q)$ on $\Gamma_{inv}$ [FRT, Ko2] since
$$\eqalign{\Delta^{{\circ}{\circ}}(S_A(\t ur)\otimes\t st) &=
          \sum_{k,l} S_A(\t kr)\otimes\t sl \otimes S_A(\t uk)\otimes\t
lt\,,\cr
          \varepsilon^{{\circ}{\circ}}(S_A(\t ur)\otimes\t st) &=
          \delta^{us}_{\ \ rt}\cr}
\eqno(3.13)
$$
where $\Delta^{{\circ}{\circ}}$ and $\varepsilon^{{\circ}{\circ}}$ are the
(dual) comultiplication and counit respectively of $D(A_q)$ according to
Theorem 2 in [CEJSZ]. Since the spaces $\fX$ and $\Gamma_{inv}$ are dual
(and finite dimensional) the map $\Phi^\circ_\Gamma$ in eq.$\,\,$(3.12)
induces a left action of the quasitriangular Hopf algebra
$\left(D(U_q(\fg)), {{\cal R}^\circ}\right)$ on $\fX$.
$$\ad^\circ:=(\,.\,)\circ\Phi^\circ_\Gamma:D(U_q(\fg))\otimes\fX\to\fX\,.
  \eqno(3.14)
$$
{}From the definition it is obvious that
$$\ad = \ad^\circ \circ (\Delta_U \otimes \id_\fX )\,.\eqno(3.15)$$
To calculate the braiding $\Psi_{0\,\fX\fX}$ (w.r.t. $D(U_q(\fg))$) one
observes
that
$$\eqalign{&<\Psi_{0\,\fX\fX}(x\otimes y), \eta\otimes\zeta >\cr
         = &< x\otimes y, ({{\cal R}^\circ}\otimes \id_\Gamma
           \otimes\id_\Gamma )\,(\Phi^\circ_\Gamma \otimes \Phi^\circ_\Gamma)\,
           (\eta \otimes \zeta)>\cr
         = &< x\otimes y, \Psi_{0\,\fX\fX}^t (\eta\otimes \zeta)>\,.\cr}
\eqno(3.16)
$$
where $x\otimes y \in \fX\otimes\fX$ and $\eta\otimes\zeta\in\Gamma_{inv}
\otimes\Gamma_{inv}$. For the basis elements
$\eta_a^{\ b}\otimes\eta_c^{\ d}\in \Gamma_{inv}\otimes\Gamma_{inv}$
one obtains
$$\Psi_{0\,\fX\fX}^t (\eta_a^{\ b}\otimes\eta_c^{\ d}) =
  \sum_{t,u,v,w} \L +uc\,S_U(\L -dt)\,\left(S_A(\t wa)\t bv\right)\cdot
  \eta_u^{\ t}\otimes\eta_w^{\ v}
  \eqno(3.17)
$$
and comparing with [DSO] yields
$$\Psi_{0\,\fX\fX} = \sigma\,. \eqno(3.18)$$
The results of Section 2 can now be applied. Through the identification
$\fX\cong\fX_1$ and for $\alpha =1$ in (2.21) one finds that
$$(\ni^{(1)}\Psi_0)_{\fX\,\fX} = -\Psi_{0\,\fX\fX} = -\sigma \eqno(3.19)$$
in the category of $\ZZ_2$-graded $D(U_q(\fg))$-left modules. Thus according
to Proposition 2.9 and eqs.$\,\,$(3.6) and (3.19) the space $\Lambda(\fX)$ is a
braided super-Hopf algebra in the category
$({}_{D(U_q(\fg))}{\cal M}, \otimes ,\KK ,\ni^{(1)}\Psi_0)$ of $\ZZ_2$-graded
$D(U_q(\fg))$-left modules. Theorem 2.5 then yields the following
corollary.
\abs
{\bf Corollary 3.2.} $\Lambda(\fX)\rtimes_{\ni^{(1)}} D(U_q(\fg))$ is a
super-Hopf algebra with structure maps defined through the eqs.$\,\,$(2.7).\lfl
$\square$
\abs
For further argumentation we need
\abs
{\bf Lemma 3.3.} The comultiplication $\Delta_U : U_q(\fg)\to
D(U_q(\fg))$ is an injective Hopf algebra homomorphism.
\abs
{\it Proof.} Since $U_q(\fg)$ is a Hopf algebra, $\Delta_U$ is an injective
$\CC$-vector space homomorphism. With the help of (3.11) it is
straightforward to verify that
$$\eqalign{\Delta_U\circ m_U &= m^\circ\circ(\Delta_U\otimes\Delta_U)\,,\cr
           \Delta_U\circ\eta_U &= \eta^\circ\,,\cr
     (\Delta_U\otimes\Delta_U)\circ\Delta_U &= \Delta^\circ\circ\Delta_U\,,\cr
    \varepsilon_U &= \varepsilon^\circ \circ \Delta_U\,,\cr
  \Delta_U\circ S_U &= S^\circ\circ \Delta_U\,.\cr}
\eqno(3.20)
$$
This concludes the proof.\lfl $\square$
\abs
The injective linear mapping $\phi := (\id_\Lambda\otimes\Delta_U) :
\Lambda (\fX)\otimes U_q(\fg) \to \Lambda (\fX)\otimes D(U_q(\fg))$
enables us to formulate
\abs
{\bf Theorem 3.4.} $\phi(\Lambda (\fX)\otimes U_q(\fg))$ is a super-Hopf
subalgebra of $\Lambda (\fX)\rtimes_{\ni^{(1)}} D(U_q(\fg))$ and thus
$\Lambda (\fX )\otimes U_q(\fg)$ becomes a super-Hopf algebra
through the mappings
$$\eqalign{m_K &:= \phi^{-1}\circ \hat m\circ (\phi\otimes\phi)\,,\cr
           \eta_K &:= \phi^{-1}\circ\hat\eta\,,\cr
   \Delta_K &:= (\phi^{-1}\otimes\phi^{-1})\circ\hat\Delta\circ\phi\,,\cr
           \varepsilon_K &:= \hat\varepsilon\circ\phi\,,\cr
           S_K &:= \phi^{-1}\circ\hat S\circ\phi\,.\cr}
\eqno(3.21)
$$
The algebra $(\Lambda (\fX)\otimes U_q(\fg),\, m_K,\,\eta_K)$
coincides with the algebra $K(q,\fg )$. Therefore the space
$(K(q,\fg),\,\Delta_K,\,\varepsilon_K,\,S_K)$ is a super-Hopf algebra.
\abs
{\it Proof.} $L:= \phi(\Lambda (\fX)\otimes U_q(\fg))$ is $\NN_0$-graded w.r.t.
the grading in $\Lambda (\fX)=\bigoplus_{j=0}^\infty \Lambda_i (\fX)$. If one
can show that $\hat m (L\otimes L)\subset L\,,\
\hat\Delta (L) \subset L\otimes L\,,\ \hat S (L) \subset L$
the proof is done. Some calculation yields
$$\eqalign{&\hat m ((\lambda\otimes u)\otimes (\mu \otimes v))
     = \sum \lambda\cdot (u_{(1)}\act\mu)\otimes \Delta_U (u_{(2)}v)\,,\cr
     &\hat\Delta (\lambda\otimes u) = \sum \lambda_{(1)}\otimes
     \Delta_U ({\cal R}_1^{-1}{\cal R}_{2'}u_{(1)})\otimes
     ({\cal R}_2^{-1}\otimes {\cal R}_{1'})\act^{\!\!\circ} \lambda_{(2)}
     \otimes \Delta_U(u_{(2)})\,,\cr
     &\hat S (\lambda\otimes u) = \sum \left(\Delta_U\,S_U({\cal R}_1^{-1}
     {\cal R}_{2'} u)_{(1)}\,({\cal R}_2^{-1}\otimes {\cal R}_{1})\right)
   \act^{\!\!\circ} S_\Lambda (\lambda )\otimes \Delta_U\,S_U({\cal R}_1^{-1}
     {\cal R}_{2'} u)_{(2)}\cr}
\eqno(3.22)
$$
where we used the quasitriangularity of $(D(U_q(\fg)), {{\cal R}^\circ})$,
and eqs.$\,\,$(2.7) and (3.10). In the above relations
$\act $ is the action $\ad^\wedge : U_q(\fg)\otimes \Lambda (\fX) \to
\Lambda (\fX)$ and $\act^{\!\!\circ}$ is the action $\ad^{\circ\,\wedge} :
D(U_q(\fg))\otimes\Lambda (\fX)\to \Lambda (\fX)$ which are induced from
$\ad$ and $\ad^\circ $ respectively.\lfl $\square$
\abs
In the last part of this section we will show that $m_K$, $\eta_K$, $\Delta_K$,
$\varepsilon_K$, and $S_K$ from the definition in (3.21) respect the chain
complex structure induced by the derivation d.
\abs
{\bf Theorem 3.5.} If the tensor product
$K^\otimes (q,\fg) = K(q,\fg)\otimes K(q,\fg)$ is
supplied with the chain complex structure according to Proposition 3.1
then the mappings
$$\eqalign{m_K &:(K^\otimes (q,\fg), {\rm D})\to (K(q,\fg),{\rm d})\,,\cr
       \eta_K &:(\CC , 0)\to (K(q,\fg),{\rm d})\,,\cr
       \Delta_K &: (K(q,\fg), {\rm d}) \to (K^\otimes (q,\fg),{\rm D})\,,\cr
     \varepsilon_K &: (K(q,\fg),{\rm d})\to  (\CC , 0)\,,\cr
       S_K &: (K(q,\fg),{\rm d})\to (K(q,\fg),{\rm d})\cr}
\eqno(3.23)
$$
are $\ZZ_2$-graded morphisms of complexes with differentiation [CE], i.e.
the diagrams
{\input amstex
\def\diag#1#2#3#4#5#6#7#8{\CD                   
                          #1 @> #5 >> #2 \\     
                          @V #6 VV @V #8 VV \\  
                          #4 @> #7 >> #3        
                          \endCD}               

$$\gathered
{\diag {K(q,\fg)}{\CC}{\CC}{K(q,\fg)}{\varepsilon_K}{\roman
d}{\varepsilon_K}{0}
             \qquad\quad
\diag {\CC}{K(q,\fg)}{K(q,\fg)}{\CC}{\eta_K}{0}{\eta_K}{\roman d}
\quad\quad
\diag {K(q,\fg)}{K(q,\fg)}{K(q,\fg)}{K(q,\fg)}{S_K}{\roman d}{S_K}{\roman d}}
\\
{\diag {K^\otimes (q,\fg)}{K(q,\fg)}{K(q,\fg)}{K^\otimes (q,\fg)}
        {m_K}{\roman D}{m_K}{\roman d}
\quad\quad
\diag {K(q,\fg)}{K^\otimes (q,\fg)}{K^\otimes (q,\fg)}{K(q,\fg)}
      {\Delta_K}{\roman d}{\Delta_K}{\roman D}}
\endgathered\tag{D-2}$$
and} the corresponding diagrams where d, D and 0 are replaced by
$\gamma$, $\Gamma$ and $\id_{\CC}$ respectively, are commutative.
\abs
{\it Proof.} The several proofs are either very similar or very simple. We
only sketch the proof of the identity $S_K\circ {\rm d}= {\rm d}\circ S_K$.
Let $\X lk \in \fX $ be a basis element of $\fX\subset\Lambda (\fX)$.
Then using the notation of [DSO] one obtains
$S_K\,d(\X lk)= S_U(\tilde \X lk)\in U_q(\fg)\,.$ On the other side one gets
$d\,S_K(\X lk) = - \sum_{a,b} S_U(\O lakb )\,\tilde\X ba $
where $\O lakb = \L +lb\,S_U(\L -ak)$. Inserting the definition of
$\tilde\X ba$ according to eq.$\,\,$(3.2)
yields $d\,S_K(\X lk)=S_U(\tilde\X lk)$.
For $u\in U_q(\fg)\subset K(q,\fg)$ the identity
${\rm d}\,S_K(u)=S_K\,{\rm d}(u)$ is trivially valid.
Now let $a,\,b\in K(q,\fg)$ be homogeneous elements of degree
$\hat a$ and $\hat b$ respectively which obey the relations
${\rm d}\,S_K(a)=S_K\,{\rm d}(a)$ and ${\rm d}\,S_K(b)=S_K\,{\rm d}(b)$.
Then $S_K\,d(a\,b)=(-1)^{(\hat a-1)\,\hat b}S_K (b)\,d\,S_K(a)+
(-1)^{\hat a\,\hat b}d\,S_K(b)\,S_K(a)= d\,S_K(a\,b)$
where the fact has been used that $K(q,\fg)$ is a super-Hopf algebra and that d
decreases the degree by 1. Since $\fX$ and $U_q(\fg)$ generate $K(q,\fg)$
it follows that $S_K\circ {\rm d}= {\rm d}\circ S_K$.\lfl $\square$
\Abs
{\bfeins References}
\Ues
\litem {[CE]} Cartan, H., Eilenberg, S.: Homological Algebra.
   Princeton Mathematical Series, Vol. {\bf 19}, Princeton (1956).
\litem{[CEJSZ]} Chryssomalakos, C., Engeldinger, R., Jur\v co, B.,
   Schlieker, M., Zumino, B.: Complex Quantum Enveloping Algebras as
   Twisted Tensor Products. Preprint LMU-TPW 93-2 (1993).
\litem{[CSWW]} Carow-Watamura, U., Schlieker, M., Watamura, S., Weich, W.:
   Bicovariant differential calculus on quantum groups $SU_q(N)$ and $SO_q(N)$.
   Commun. Math. Phys. {\bf 142}, 605 (1991).
\litem {[DJSWZ]} Drabant, B., Jur\v co, B., Schlieker, M.,
   Weich, W., Zumino, B.: The Hopf Algebra of Vector Fields on Complex
   Quantum Groups. Lett. Math. Phys. {\bf 26}, 91 (1992).
\litem{[Dri]} Drinfel'd, V. G.: Quantum groups.
   Proceedings of the International Congress
   of Mathematicians, Berkeley 1986, 798 (1986).
\litem{[DSO]} Drabant, B., Schlieker, M., Ogievetsky, O.:
   Cohomology of Quantum Enveloping Algebras. Preprint MPI-Ph/93-57,
   LMU-TPW 1993-19 (1993). Submitted to Commun. Math. Phys.
\litem{[FRT]} Faddeev, L.D., Reshetikhin, N.Yu., Takhtajan, L.A.:
   Quantization of Lie Groups and Lie Algebras.
   Leningrad Math. J. {\bf 1}, 193 (1990).
\litem{[JS]} Joyal, A., Street, R.: Braided Monoidal Categories.
   Mathematics Reports 86008, Macquarie Univ. (1986).
\litem{[Jur]} Jur\v co, B.: Differential Calculus on Quantized Simple
   Lie Groups. Lett. Math. Phys. {\bf 22}, 177 (1991).
   {\it Lett. Math. Phys.}{\bf 22}, 177 (1991).
\litem{[Ko1]} Koornwinder, T.H.: Orthogonal Polynomials in Connection
   with Quantum Groups. NATO ASI Series {\bf C 294}, P. Nevai (ed.),
   Kluver Acad. Publishers (1990).
\litem{[Ko2]} Koornwinder, T.H.: General Compact Quantum Groups, a Tutorial.
   Preprint FWI-Report 94-06 (1994). To appear in ``Representations of Lie
   Groups and quantum Groups'', M. Picardello (ed.), Longman.
\litem{} Dijkhuizen, M.S., Koornwinder, T.H.: CQG Algebras: A direct Algebraic
   Approach to Compact Quantum Groups. Preprint CWI-Report AM-R 9401 (1994).
   To appear in Lett. Math. Phys.
\litem{[Mac]} Mac Lane, S.: Categories for the Working Mathematician.
   Graduate Texts in Mathematics {\bf 5}, Springer (1972).
\litem{[Man]} Manin, Yu.I.: Topics in Noncommutative Geometry. Princeton
   Univ. Press, Princeton (1991).
\litem{[Ma1]} Majid, S.: Braided Groups. J. Pure Appl. Algebra {\bf 86}, 187
   (1993).
\litem{[Ma2]} Majid, S.: Cross Products by Braided Groups and Bosonization.
   J. Algebra {\bf 163}, 165 (1994).
\litem{[Ma3]} Majid, S.: Quasitriangular Hopf Algebras and Yang-Baxter
   Equations. Int. J. Mod. Phys. {\bf A5} Vol. 1, 1 (1990).
\litem{[Ma4]} Majid, S.: Quantum and Braided Linear Algebra. J. Math. Phys.
   {\bf 34}, 1176 (1993).
\litem{[Ma5]} Majid, S.: Braided Momentum in the $q$-Poincar\'e Group.
   J. Math. Phys. {\bf 34}, 2045 (1993).
\litem{[Par]} Pareigis, B.: A Non-Commutative Non-Cocommutative Hopf
   Algebra in ``Nature''. J. Algebra {\bf 70}, 356 (1981).
\litem {[Res]} Reshetikhin, N.Yu.: Quantized Universal Enveloping Algebras,
   the Yang-Baxter Equation and Invariants of Links I.
   Preprint LOMI E--4--87 (1987).
\litem{[SWW]} Schlieker, M., Weich, W., Weixler, R.O.: Inhomogeneous
   Quantum Groups. Z. Phys. {\bf C 53}, 79 (1992).
\litem{[SWZ]} Schupp, P., Watts, P., Zumino, B.: Bicovariant Quantum Algebras
   and Quantum Lie Algebras. Commun. Math. Phys. {\bf 157}, 305 (1993).
\litem{[SZ]} Schlieker, M., Zumino, B.: Braided Hopf Algebras and Differential
   Calculus. Preprint LBL-35299, UCB-PTH-94/03 (1994).
\litem{[Wei]} Weixler, R.O.: Inhomogene Quantengruppen. Thesis.
   Shaker, Aachen (1994).
\litem{[Wor]} Woronowicz, S.L.: Differential Calculus on Compact Matrix
   Pseudogroups (Quantum Groups). Commun. Math. Phys. {\bf 122}, 125 (1989).
\FF
\end